\documentclass[fleqn,10pt]{olplainarticle}
\usepackage{pdflscape}
\usepackage{newfloat}
\usepackage{hyperref}
\usepackage{lineno}
\usepackage{setspace}

\hypersetup{
    colorlinks=true,
    citecolor=black,
    filecolor=black,
    linkcolor=black,
    urlcolor=black
    }

\title{A computer vision-based approach to clean seismic catalogues}

\author[1]{Michele De Solda*}
\author[3]{Camilla Rossi}
\author[2]{Sonja Gaviano}
\author[2]{Giacomo Rapagnani}
\author[2]{Emanuele Bozzi}
\author[4]{Bogdan Enescu}
\author[2]{Laura Gulia}
\author[2]{Francesco Grigoli*}

\affil[1]{Department of Earth Sciences, Sapienza University, Rome, Italy}
\affil[2]{Department of Earth Sciences, University of Pisa, Pisa, Italy}
\affil[3]{Seismix srl, Palermo, Italy}
\affil[2]{Department of Earth and Planetary Sciences, Kyoto University, Kyoto, Japan}
\affil[*]{michele.desolda@uniroma1.it, francesco.grigoli@unipi.it; These authors contributed equally to the manuscript.}


\begin{abstract}
 In recent years, seismic data analysis advancements combined with an increasing number of dense seismic networks deployed worldwide, have contributed to the creation of massive seismic catalogs, significantly lowering their magnitude of completeness. However, large automated catalogs are typically released without systematic quality control, and may contain spurious detections, mislocations, or inconsistent magnitudes. In challenging scenarios, such as microseismic monitoring applications, where weak and closely spaced events often overlap in time, pick-based detection and location approaches often fail to reliably associate phases. This leads to missed detections or degraded location accuracy producing seismic catalogues polluted with false or mislocated events. To address this limitation, we present a computer vision–based workflow that integrates waveform-based seismic location methods with deep learning image classification to discriminate real seismic events from noise directly from coherence matrices. These matrices, computed via waveform stacking, exhibit distinct patterns for real events (single, focused maxima) versus noise (blurred, incoherent patterns) hence the problem of cleaning seismic catalogues can be solved as a binary image classification problem. In addition, the robustness of waveform-based location methods allows to obtain an increased resolution in the location of seismic events. Another advantage of this approach is that the training of neural networks can be based entirely on synthetic data, using stochastic modeling to generate noise recordings with realistic spectral properties and synthetic waveforms based on pre-built Green’s functions databases. This synthetic-based training removes the need for large labeled datasets, enabling rapid deployment in newly instrumented areas. We validate our workflow using the publicly available COSEISMIQ dataset from the Hengill geothermal area, in Iceland. We finally demonstrate that our method removes spurious detections, flags mislocated events, and relocates true events with improved accuracy, yielding an enhanced catalog that shows better features than the double-difference seismic catalog in the area, while retaining a significantly larger number of events. This approach provides a robust, scalable solution for cleaning and enriching seismic catalogs, supporting more reliable seismotectonic interpretation and seismic hazard assessment.

\end{abstract}

\begin{document}
\flushbottom
\maketitle
\thispagestyle{empty}

\doublespacing

\section*{Declaration of Competing Interests}
The authors acknowledge there are no conflicts of interest recorded.

\section*{Introduction}
The growing number of dense seismic networks worldwide is resulting in the generation of increasingly large volumes of seismological data (\cite{Kong2019}). The total volume collected annually by global seismographic networks is now on the order of petabytes (1 PB = 1024 TB) \cite{arrowsmith_big_2022}. This data explosion, combined with increasingly powerful computational capabilities, has enabled the development and application of sophisticated data analysis methods that can digest massive seismicity datasets and have the potential to outperform standard seismic processing routines and even experienced seismologists (\cite{Mous-Ber2022}).

Among these, deep learning (DL) techniques are particularly well-suited for addressing the challenges posed by the exponential growth of seismic datasets (\cite{bergen_machine_2019}). DL-based frameworks have been used to generate massive seismicity catalogues \cite[e.g.,][]{tan2021}. In particular, DL pickers \cite[e.g.,][]{phasenet, CRED, eqtransformer,  GPD, DKPN, DPP, LPPN, Woollam_2019, CPIC, ppknet, {seisbench}} and associators \cite[e.g.,][\cite{phaselink_2019}]{genie, gamma, malmi_2022, Si_2024, PyOcto, Woollam_2020, McBrearty_2019} can significantly increase the number of detected events, lowering the magnitude of completeness of seismicity catalogs by several units. Such catalogs may contain an order of magnitude more events than the catalogs routinely produced by local earthquake monitoring agencies (\cite{tan2021, chiaraluce2022}). However, the performance of DL models on new datasets can be subject to model training, model generalization ability, and choice of threshold parameters (\cite{Munchmeyer2022}, \cite{Lim2025}). As a result, the problem of reliability of these automatically generated catalogues is thus of paramount importance, especially when applying pre-trained models with unseen challenging datasets like, for instance, seismic sequences characterized by a large number of weak earthquakes overlapping each other or with short inter-event times. In such cases even the most sophisticated methods may struggle to correctly detect, identify and associate the picks of specific seismic phases (\cite{Grigoli2018GJI}). Due to the inability to manually check each single event in such large catalogues, potential inconsistencies in event location and magnitude can have severe implications on the estimation of spatio-temporal variation of b-value and on the performance of earthquake forecasting models used for risk assessment (\cite{Herr_Marz_2020}; \cite{mancini}). Artifacts in seismic catalogues such as spatial-temporal complexity in completeness, shifts or stretches of the magnitude scale, and magnitude-dependent location uncertainty may have pernicious effects in the estimation of parameters used in statistical seismology, highlighting the importance of producing seismic catalogues that are not only rich in terms of the number of events but also clean and reliable. New ML-based catalogues are also impacted by these problems (\cite{Herr_Marz_2020}). Careful data mining can reveal many of these problems, but so far, these approaches are based largely on visual inspections of diagnostic plots, making their application to large catalogues infeasible. In this study, we propose an automated framework that simultaneously cleans seismic catalogues (by discriminating real earthquakes from noise) and enhances them by relocating the events using waveform-based methods. More specifically, we combine waveform stacking location techniques \cite[e.g.,][]{ Grigoli2013GRL, bbtrack, Gharti2010, KaoSSA, Li_2020, stanek} with computer vision methods commonly used for image classification \cite[e.g.,]{resnet,LeCun, Zhao_2024}. We validate the proposed workflow using the publicly available dataset from the COSEISMIQ project (\cite{grigoli2022}), which consists of two years of continuous seismic recordings from a temporary microseismic network in the Hengill geothermal area, Iceland. The dataset also includes reference seismicity catalogues, which are used both to generate synthetic training data and to benchmark the performance of our framework.

\section*{Dataset}
\begin{figure}[ht]
\centering
\includegraphics[width=1.0\linewidth]{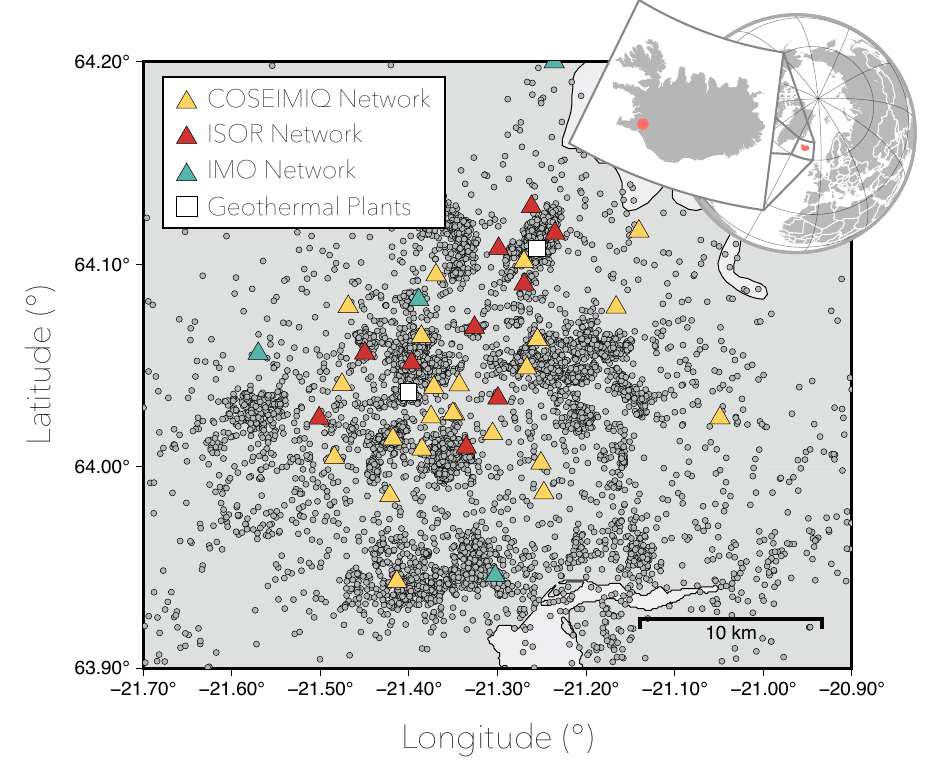}
\caption{Map of the Hengill geothermal field. Grey dots represent the location of the events of the low-quality (LQ) catalog. Triangle markers represent seismic stations belonging to: regional network of the IMO (red, IV), permanent stations of the ISOR (blue, OR), and the temporary COSEISMIQ networks (yellow, 2C, 4Q). White squares are the the two power plants Hellisheidi and Nesjavellir.   }
\label{fig1}
\end{figure}

The data we used to develop and test our methodology comes from microseismic monitoring operations carried out in the Hengill geothermal field (in Iceland) during the EU project (COSEISMIQ) (Control SEISmicity and Manage Induced Earthquakes) \cite{grigoli2022}. The Hengill region is one of the most seismically active places on Earth, with several thousand earthquakes recorded yearly. Here, natural and induced seismicity coexist due to the presence of two large geothermal power plants (Hellisheidi and Nesjavellir) (Figure \ref{fig1}). This region is currently monitored by 8 seismic stations of the regional network of the Icelandic Meteorological Office (IMO, network code VI) and by 10 permanent seismic stations managed by Icelandic Geosurvey (ISOR, network code OR). Within COSEISMIQ, the permanent seismic network has been densified by ETH-Zurich and GFZ-Potsdam with about 40 temporary stations (both short-period and broadband sensors with network codes 2C and 4Q) that operated between December 2018 and August 2021 (Figure \ref{fig1}). Broadband and short-period seismic stations of the 2C, OR, and 4Q networks operate at a sampling rate of 200 Hz while the stations of the VI network run at 100 Hz. 

The COSEISMIQ project dataset consists of publicly available continuous seismic waveforms and seismicity catalogs of different quality \cite{grigoli2022}. Raw continuous waveforms, station metadata and catalogs are openly distributed through FDSN webservices at the European Integrated Data Archive (EIDA) \cite{grigoli2022}. Seismicity in the Hengill region is characterized by a large number of weak earthquakes overlapping each other or with short inter-event times (Figure \ref{fig2}.a). In these cases, pick-based detection and location methods may struggle to correctly assign picks to phases and events, and errors can lead to missed detections and/or reduced location resolution. Similar problems are caused by the numerous off-network earthquakes (with $t_{s}-t_{p}>10-15$ s) occurring in the adjacent regions (Figure \ref{fig2}.b). During the COSEISMIQ experiment, three automated seismicity catalogs (magnitude of completeness around 0.5) of different quality levels were generated. To be included in any catalog, an event must have at least 10 associated phase picks. Furthermore, each catalog only contains the events located within the following geographical region: Latitude (North) from 63.9 to 64.2; and Longitude (East) from -21.7 to -20.9 degrees (\cite{grigoli2022}). Each catalog is defined using a scoring metric that combines azimuthal gap, phase count, RMS residuals, source–station distances, and pick residuals into a single negative-valued score (see \cite{grigoli2022} for details). The high-quality (HQ), medium-quality (MQ), and low-quality (LQ) catalogs, are separated by a subclassification of the score: $S \geq -1$, $-5 < S < -1$, and $S \leq -5$, respectively. Applying an increasingly strict score thresholds reduces the catalog sizes from approximately 12500 events (LQ), to 9900 (MQ), and finally to 8700 events (HQ). A further refined seismic catalog (HQDD), derived from the (HQ), has been obtained by applying the Double-Difference location method (\cite{Waldhauserdd}) which provides a higher resolution location of seismic events.  While the HQDD and HQ catalogs offer the most reliable locations and are suitable for detailed seismotectonic analysis, the MQ catalog supports broader statistical studies. On the other hand, the LQ seismicity catalog represented in Figure\ref{fig1} should be used with caution, as it may contain inaccurate locations and could be polluted by false or poorly constrained events (\cite{grigoli2022}). In this work, we will use the HQDD catalog as a reference to compare our enhanced seismic catalog obtained using the LQ catalog as input of our framework.

\begin{figure}[ht]
\centering
\includegraphics[width=1.0\linewidth]{figure_2.pdf}
\caption{Seismic signals of 4 closely-spaced events occurring in the studied area in a time-span of less than 30 s (a). Seismic signal of an event located outside the area of interest (b). Example of seismic waveforms dominated by microseismic noise in the 5-12 s period band before (c) and after high-pass filtering (d).}
\label{fig2}
\end{figure}

Another problem affecting the COSEISMIQ dataset is related to the strong ambient noise that characterizes Iceland. This high level of noise affecting the seismic waveforms, mainly in the 3–20 s period band (primary and secondary microseims), is capable to mask earthquakes up to magnitude 1.5 (\cite{grigoli2022}). To mitigate the impact of this noise, the waveforms were high-pass filtered using a second-order Butterworth filter with a corner frequency of 2 Hz. The effect of this filtering process is illustrated for a representative noise recording in Figure \ref{fig2}. The scope of this work is to clean and enhance the LQ seismic catalogue and compare it with the HDQQ.

\section*{Methodology}
Our workflow for cleaning and enhancing seismic catalogs combines deep learning–based image classification techniques with waveform-based coherence analysis to discriminate seismic events of interest from noise (e.g., noise recordings or off-network earthquakes). Waveform-based methods have gained significant attention due to their ability to detect and locate earthquakes without relying on traditional phase picking and association procedures \cite[]{ Grigoli2013GRL, bbtrack, Gharti2010, KaoSSA, Li_2020, stanek}. Unlike conventional pick-based approaches, which rely on station-wise data, these methods exploit the information from the entire seismic network to produce multi-dimensional coherence matrices (i.e., images) that exhibit a maximum at the hypocentral coordinates of the seismic event. The image patterns within these coherence matrices can be used to distinguish true seismic events from false detections (i.e., noise or teleseismic events). Typically, coherence matrices for real (in-network) earthquakes display a single, well-focused maximum, while those generated from pure noise (or from off-network earthquakes) tend to show blurred patterns with low coherence values or multiple poorly focused maxima. Deep learning algorithms are well-suited for classifying such images. The objective of this work is to develop a framework that locates seismic events and classifies earthquake signals against noise. The core methodological innovation is the use of coherence matrices, obtained from waveform stacking procedures, as structured input to a Convolutional Neural Network (CNN), to solve a binary image classification problem. Due to the limited availability of labeled events in newly deployed networks, the main advantage of this approach is that CNN can be trained efficiently using synthetic data.

\subsection*{Synthetic Data Generation}
\begin{figure}[ht]
\centering
\includegraphics[width=1.0\linewidth]{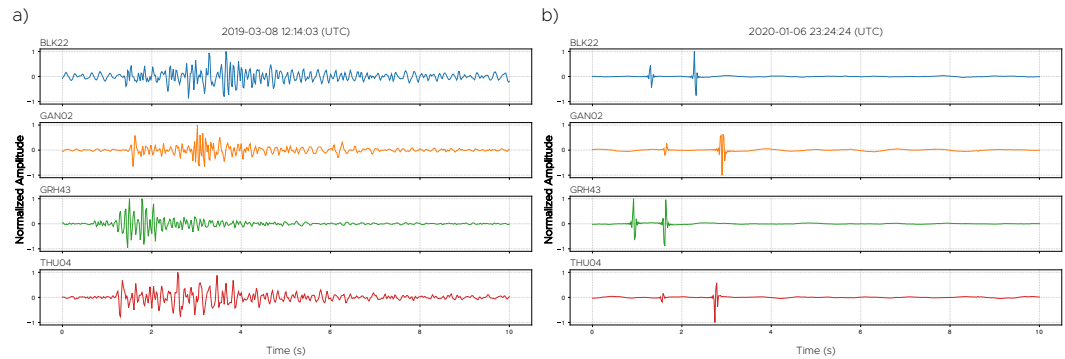}
\caption{Comparison of seismic event waveforms: real seismic waveforms (a) and synthetic ones (b)}
\label{fig3}
\end{figure}

In seismology, the performance of supervised deep learning models for earthquake detection is highly dependent on the quality of the training dataset.  When working with newly deployed seismic networks the lack of human-labeled data makes the application of DL methods challenging, and pre-trained models are often not suitable for the analysis. A potential solution would be the use of synthetic training datasets, but the inherent difficulty in reproducing synthetic waveforms with the complex characteristics of real seismograms is the main limitation of this approach (Figure \ref{fig3}). One of the goals of this study is to overcome this limitation by transforming both real and synthetic seismic waveforms into a representation that reduces these differences, making them easier to compare and more similar in appearance. The framework for the synthetics generation implemented in this work requires two main ingredients: 1) modeling of synthetic waveforms for seismic events, given source locations, focal mechanisms and a velocity model characteristic of the area and 2) stochastic simulation of noise waveforms. We first focus on the modeling of synthetic waveforms. The synthetic waveforms of microseismic events used for this study are created starting from a store of pre-calculated Green Functions (GF) (\cite{Heimann_2019}). The GF store is computed using an algorithm based on the reflectivity method (\cite{Fuchs_1971, Wang_1999}). The store provides the P and S arrivals for a 1D Iceland velocity model (\cite{Tryggvason}). Event hypocenters were uniformly sampled within the study volume (latitudes: from 63.9 to 64.2 degrees; longitudes: from -21.7 to -20.9 degrees; and depth: from 1 to 15 km), and magnitudes Mw distributed uniformly between 0.5 (the magnitude of completeness of the catalogues) and 2.0. Seismic sources were simulated by considering a double-couple focal mechanism, with strike, dip and rake randomly extracted from a uniform distribution and sampled as proposed by (\cite{Nooshiri_2021}). The synthetic waveforms are then convolved with the real station instrument response, and the actual geographic coordinates of the network stations are used to replicate realistic source–receiver configurations.

Noise waveforms are generated through a stochastic procedure, starting from random sampling of real noise segments from the COSEISMIQ network in the period 2020‑03‑01 - 2020‑03‑07. Noise may originate from natural sources (e.g., wind, oceanic microseisms) or anthropogenic activities (e.g., traffic, machinery). Additionally, stations may introduce their own noise due to intrinsic self-noise, site conditions, installation quality, or equipment malfunctions. We define noise as any ground motion recorded by the seismometer that is not generated by an earthquake, whether natural or induced, inside our control volume. In our modeling, noise is treated as a realization of a stochastic process that is stationary over the maximum time span of a seismic event recorded by the network (around 30 s). We intentionally retained noise segments containing amplitude spikes and clipped signals. This choice ensures that the classifier is exposed to challenging real-world noise patterns that could mislead threshold-based single-station detectors. 
For randomly sampled time in the selected period we extract, for all the available seismometers, real noise segments of the same length. We move in the frequency domain through the FFT and calculate the amplitude and phase spectra. We replace the original phase spectrum with random phases uniformly distributed between $-\pi$ and $\pi$, preserving the amplitude spectrum. We then move back to the time domain by applying the inverse FFT. In this way we are able to obtain a synthetic noise trace with realistic spectral content, which can be used for training our Deep Learning classifier. The main advantage of this approach is that it allows creation of synthetic noise traces only needing a few days of real continuous seismic recordings.

\subsection*{Coherence Matrix Computation}
\begin{figure}[ht]
\centering
\includegraphics[width=1.0\linewidth]{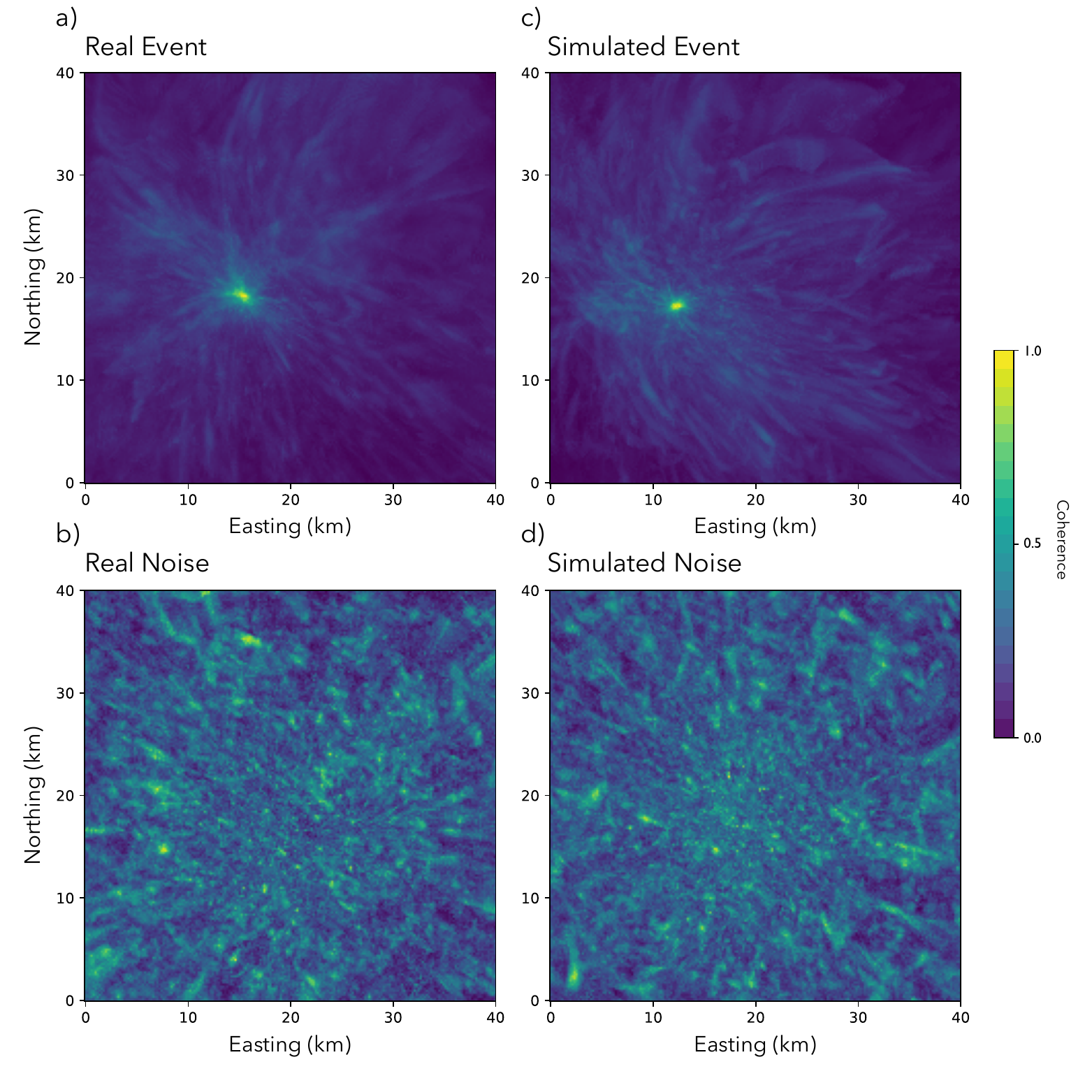}
\caption{Comparison of matrices obtained from processing real (a, c) and synthetic (b, d) waveforms. Panels (a) and (b) correspond to waveforms containing event signals, while panels (c) and (d) correspond to waveforms containing only noise.}
\label{fig4}
\end{figure}

We applied a waveform stacking-based location algorithm (\cite{Grigoli2013GRL}) to process both synthetic and real seismic waveforms. This method systematically scans a predefined 3D grid of potential source locations and origin times. For each grid point, theoretical P- and S-wave travel times are computed and used to time-align the waveforms recorded across the network. The aligned signals are then stacked to enhance coherent energy, and a coherence value is assigned to each grid point, resulting in a 4D volume of coherence values (space and time). These coherence matrices act as spatial and temporal snapshots of the stacked seismic energy, highlighting candidate source locations. The technical details of how coherence matrices are generated through waveform stacking can be found in (\cite{Grigoli2013GRL}). We perform a simple dimension reduction by taking the maximum coherence along depth and time dimensions to obtain a 2D matrix. Essentially, we transform raw seismic waveforms into coherence images in the X-Y plane at a given time, where the similarity between real and synthetic waveforms becomes more apparent and easier to compare (see Figure \ref{fig4}). Also for noise the coherence matrices generated from real data and the one generated from synthetic recordings show comparable patterns. As illustrated in Figure \ref{fig4}, real seismic events generate sharply peaked, well-focused coherence patterns, while noise or poorly located events yield diffuse, incoherent structures. This transformation offers a powerful and intuitive method for distinguishing between true seismic events and background noise, as well as for evaluating location quality.

\subsection*{Deep learning Classification}

\begin{figure}[ht]
\centering
\includegraphics[width=1.0\linewidth]{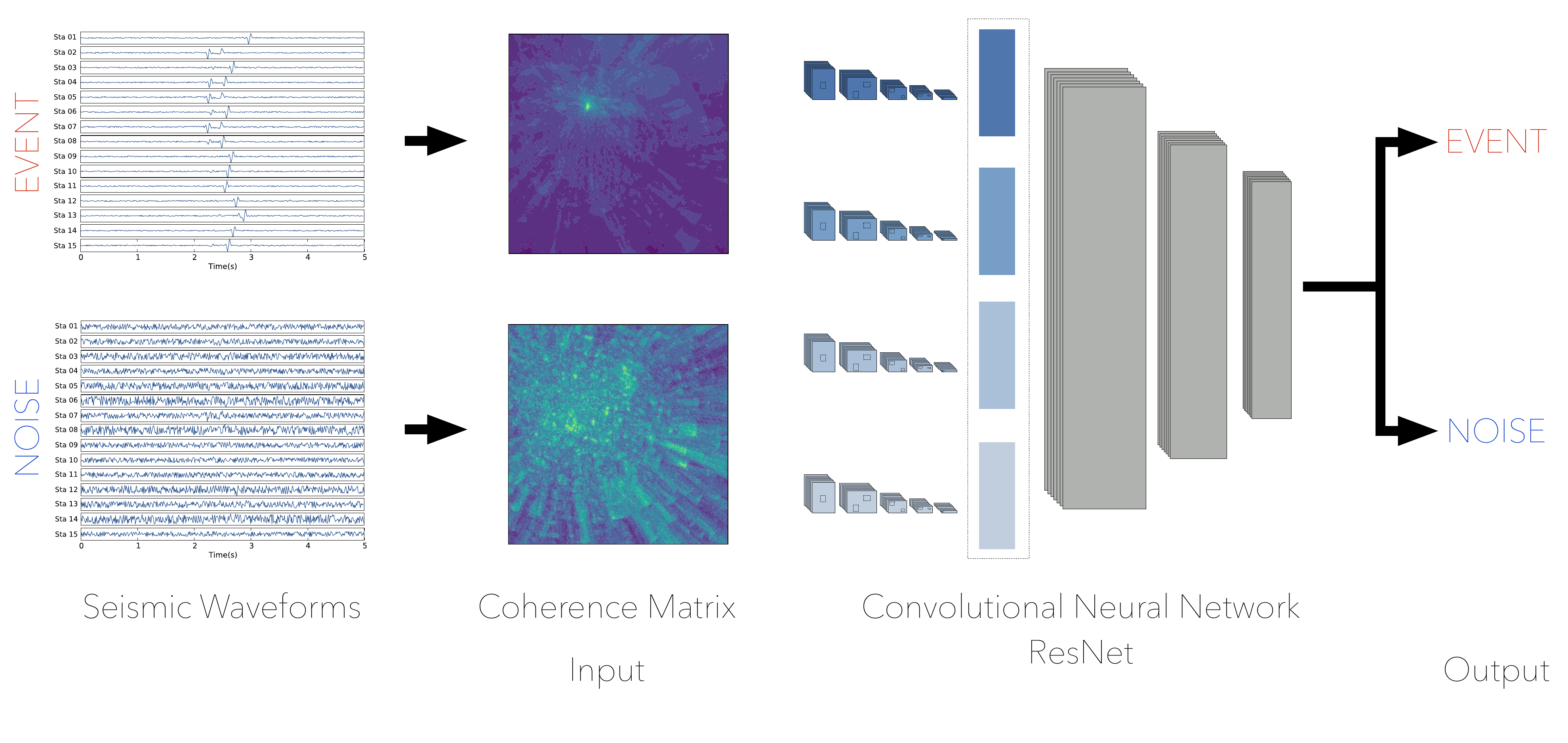}
\caption{Workflow sketch: seismic waveforms at multiple stations are transformed into coherence matrices, which are fed into the Resnet. The neural network classifies the input as belonging to an event ($score=1$) or noise ($score=0$). }
\label{fig5}
\end{figure}

We approach the problem of cleaning and enhance large, automatically generated, seismicity catalogues as a problem of image classification of coherence matrices. The classification of noise vs events coherence matrices is a quite simple task for a state-of-the-art deep learning algorithm. Such algorithms have to simply learn to distinguish an image with one well-focused maximum in the X-Y plane (Figure \ref{fig4}.a and Figure \ref{fig4}.c both for real and synthetic events) against an image with many blurred maxima and without a clear pattern (Figure \ref{fig4}.b and Figure \ref{fig4}.d both for real and simulated noise). A Convolutional Neural Network (CNN) architecture was selected, as it constitutes the standard and most appropriate approach for image classification tasks, consistent with the principles of simplicity and adaptability adopted throughout this work. Given the many solutions at our disposal, we have chosen to implement a ResNet (\cite{resnet}).
A ResNet is a CNN in which convolutional layers are combined with modules called Residual Units (RU). Each RU is a stack of two convolutional layers followed by a Batch Normalization and a ReLu activation. The key characteristic of a RU is its skip connection through which the input is added to the output. This design forces the network to model the residual relative to the identity mapping, a principle known as residual learning. Consequences are the speeding up of the training, and the possibility of training deeper networks (\cite{resnet}, \cite{hands_on}). \\
Our ResNet was implemented using the Keras functional API of Tensorflow (\cite{keras}, \cite{tensorflow}). The final architecture consists of an initial convolutional layer that takes as input coherence images (i.e matrices), followed by a max-pooling layer and a stack of 4 RUs  and a fully connected layer that outputs a prediction probability (classification score) corresponding to the classification in \textit{noise} (score=0) and \textit{event} (score=1). 
For the training, we chose the Adam optmizer, a variant of stochastic gradient descent, and the Binary cross-entropy as cost function. A schematic sketch illustrating how our proposed workflow works is shown in Figure \ref{fig5}.

\subsection*{Training and Test}

\begin{figure}[ht]
\centering
\includegraphics[width=1.0\linewidth]{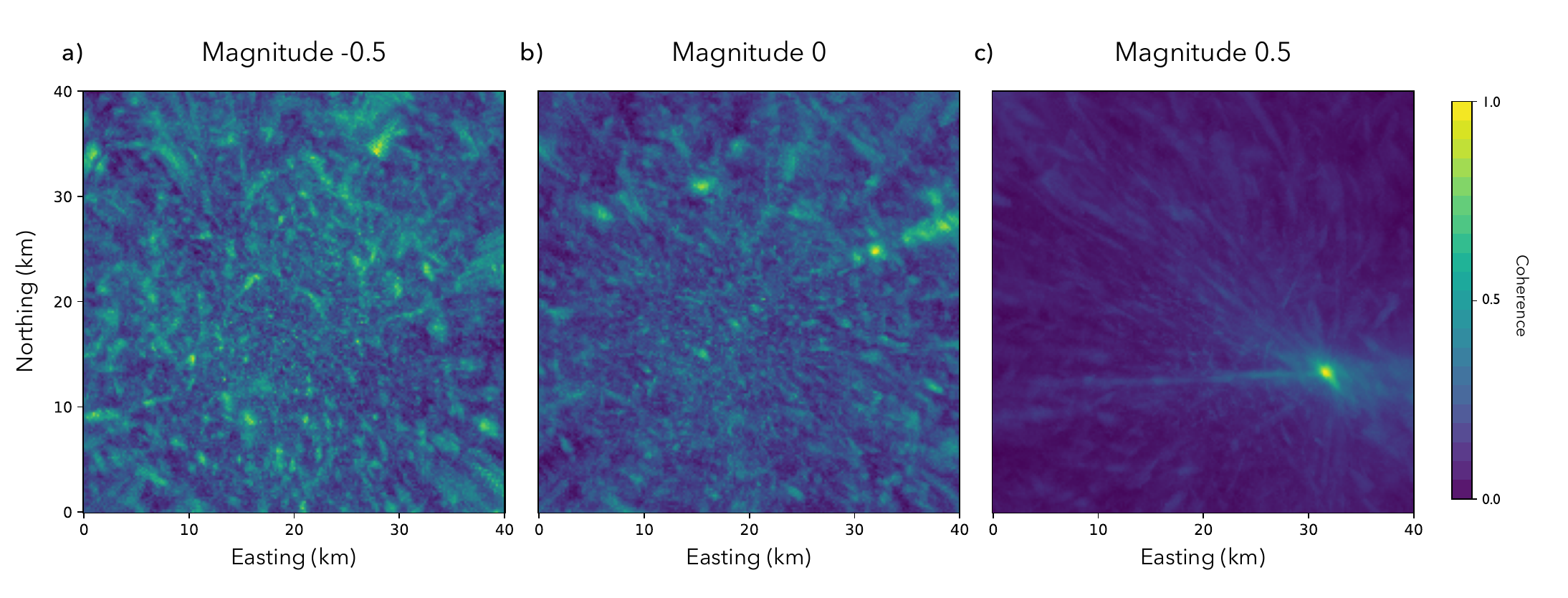}
\caption{Synthetic coherence matrices for events of different magnitudes. Below magnitude 0 the coherence matrix is similar to the one of pure noise.}
\label{fig6}
\end{figure}

To perform the training of the CNN we used a synthetic dataset composed of 4984 labeled coherence matrices, evenly splitted between noise and event ones. 90\% of the dataset was randomly selected for training and validation, while the remaining 10\% was reserved for testing. Training was performed by feeding the network with mini-batches of 32 instances over 10 epochs with early stopping based on validation loss, in order to prevent overfitting. After 5 epochs, with a validation accuracy of 99\%. When evaluated the performances on the test dataset, the trained network achieved an accuracy of 99\%. \\
Additional tests have been carried out on both synthetic and real datasets from the COSEISMIQ network. First, we evaluated the CNN's classification performance on an independent dataset of 998 coherence matrices of synthetic seismic events. The synthetic earthquakes were generated with magnitude Mw ranging from -2 to 2. The earthquake signals were then superimposed with synthetic noise generated following the approach described in the previous sections. Finally, we applied waveform stacking and transformed the seismic traces into coherence matrices as described previously. Due to the high ambient noise level characterizing the area (Figure \ref{fig2}.c and \ref{fig2}.d, for events below Mw 0 (although depending on how far is the event from the closes stations these patterns can also affect events up to magnitude 0.3) the signal-to-noise ratio tend to be generally smaller than one at almost all stations. This is also shown by Figure \ref{fig6}, showing the synthetic coherence matrices for events of magnitudes -0.5, 0 and 0.5 respectively. 

\begin{figure}[ht]
\centering
\includegraphics[width=1.0\linewidth]{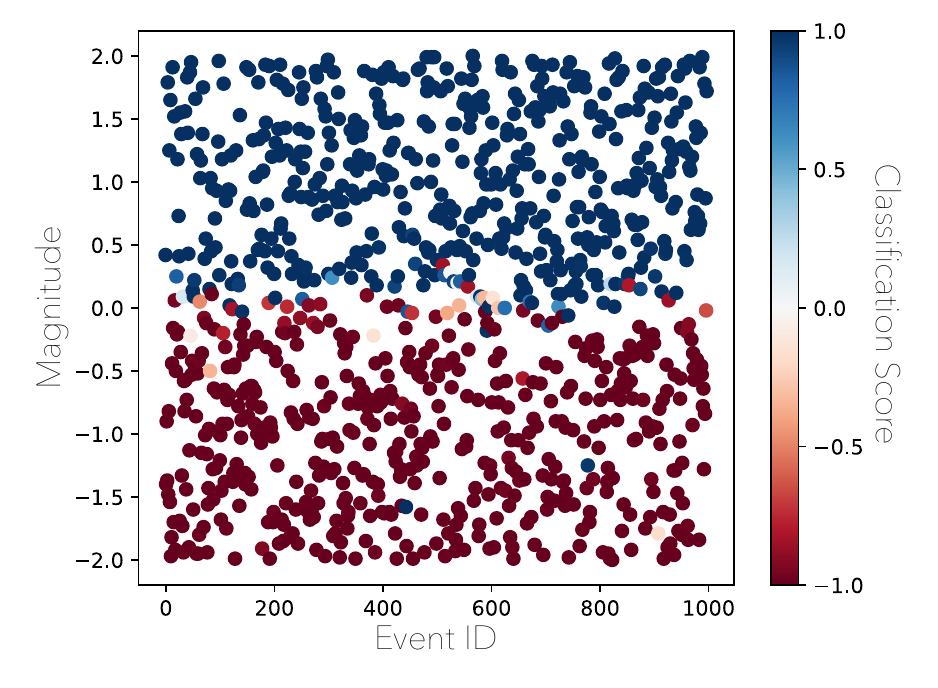}
\caption{Classification matrix showing classification performance on the synthetic test dataset. Scores closer to one indicate high confidence in classifying coherence matrices as seismic events. Here we modified the values of the score to range between (-1,1) instead of (0,1) for visualization purposes.}
\label{fig7}
\end{figure}

For events close to Magnitude 0, the signal is buried within the noise and the corresponding coherence matrices, from a qualitative analysis, tend to be indistinguishable from those generated from pure noise recordings. Figure \ref{fig7} shows the classification results of the CNN applied to this dataset. Here, for plotting purposes, the classification score ranges between -1 (noise classification) and 1 (event classification). Figure \ref{fig7} also highlights that the detector behaves as expected: events with higher magnitudes (above Magnitude 0.3) tend to receive classification scores closer to one, indicating confident identification as real events, while those with lower magnitudes (roughly below Magnitude 0.) are characterized by scores close to zero or negative, suggesting they are more likely to be classified as noise. Low-magnitude events (below Magnitude 0.3) exhibit signal-to-noise ratios close to or below one for most of the seismic stations. Consequently, their coherence matrices are visually and statistically similar to those generated from pure noise, making the classification process challenging. The CNN’s inability to correctly classify these weak events is not an indication of poor performance of the model, but rather a realistic limitation imposed by the data’s physical properties, related to the fact that very weak events will inevitably remain buried in the background noise (as shown by the coherence matrices in Figure \ref{fig5}). Figure \ref{fig7} thus serves as both a validation of the CNN’s training on synthetic coherence matrices and an illustration of the trade-offs inherent in automated seismic catalog cleaning.

\section*{Application to the COSEISMIQ data}
\begin{figure}[ht]
\centering
\includegraphics[width=1.0\linewidth]{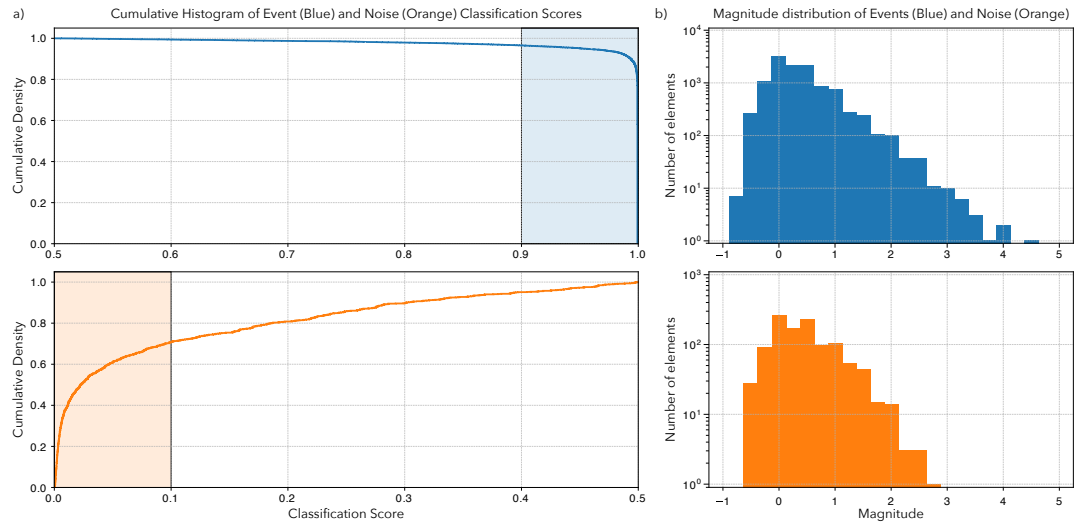}
\caption{Panels on the left show the cumulative distribution of classification scores assigned by the CNN: seismic events have a score closer to 1, while noise instances have a score near 0. Panels on the right show the magnitude distribution of classified events and noise.}
\label{fig8}
\end{figure}

To evaluate the performance of our framework with a real case, we applied the trained CNN to coherence matrices obtained from the seismic waveforms of earthquakes recorded by the COSEISMIQ network and listed in the LQ catalog. This catalog consists of 12,374 seismic events (the original count was 12,423, but 49 events were duplicates referring to the same events) and it is the largest automated seismic catalog produced during the COSEISMIQ project. The only quality constraint that \cite{grigoli2022} has applied in compiling this catalog was to set a minimum of 10 associated phases per event, with locations restricted to the region of interest defined in the previous sections. By adding further quality constraint the number of events reduces to 8691 events for the HQDD catalog, which represents the best available automatic catalogue produced by the COSEISMIQ project. In order to validate our workflow, we first applied LOKI (\cite{Grigoli2013GRL}) to the events of the LQ catalog, starting from the initial hypothesis that all the objects in this catalog are true seismic events. For each event, LOKI generates a coherence matrix that serve as input to the CNN classifier, which evaluates whether the event corresponds to a true seismic signal or to noise (considering noise also the events that are not occurring within the region of interest). The CNN classifier returns a score between 0 and 1, with values near 0 indicating noise and values near 1 indicating a true seismic event. After this process, the CNN classification scores exhibit a bimodal distribution (Figure \ref{fig8}.a), clearly separating real events (with scores approaching 1) from noise (scores near 0). This demonstrates the ability of the network to robustly distinguish coherent seismic energy from background noise using spatial patterns in coherence matrices. A comparison of the magnitude distributions for each class (Figure \ref{fig8}.b) shows that the majority of noise-classified instances occur at lower magnitudes (mostly below 0.5), where signal-to-noise ratio of seismic events is inherently low and coherence images become indistinguishable from those that would be generated by pure noise. The results of the classification is summarized in Table \ref{table:results}:

\begin{table}[ht]
\centering
\begin{tabular}{lll}
\hline
\textbf{Description} & \textbf{Number of Instances} & \textbf{Magnitude Range} \\
\hline
Total number of instances classified & 12374 & from  -0.9 to 4.5  \\
Number of event-classified instances & 11272 & from  -0.9 to 4.5  \\
Number of noise-classified instances &  1102 & from  -0.7 to 2.7  \\
\hline
\end{tabular}
\caption{Results of the classification of the coherence matrices of the events of the LQ catalog of the COSEISMIQ project.}
\label{table:results}
\end{table}

Figure \ref{fig8}.b also shows noise-classified instances with magnitudes reaching up to 2.5 in the original catalog, raising the question of whether these events are truly noise or simply misclassified. To address this issue, we manually inspected all 1102 noise-classified instances, finding that 965 were correctly classified as noise, while the remaining 137 were misclassified and represent seismic events of interest. Most misclassified events have magnitudes below 0.5 and are characterized by low signal-to-noise ratios, with the earthquake signal detectable at only a few stations, resulting in coherence matrices that closely resemble those of noise events (as in Figure \ref{fig6}). Only five cases exhibited magnitudes between 0.6 and 0.9; in these instances, the misclassification is due to multiple partially overlapping events, which generated inconsistent coherence matrices. Since the classification score is not one-hot but a continuous value Figure \ref{fig8}.b also shows that approximately 75\% of noise-classified instances have scores below 0.1, beyond this threshold the derivative of the cumulative distribution remains more or less constant. As a rule of thumb, beyond the score at which the derivative of the cumulative distribution stabilizes, it is useful to manually inspect the classified instances. 

\begin{figure}[ht]
\centering
\includegraphics[width=1.0\linewidth]{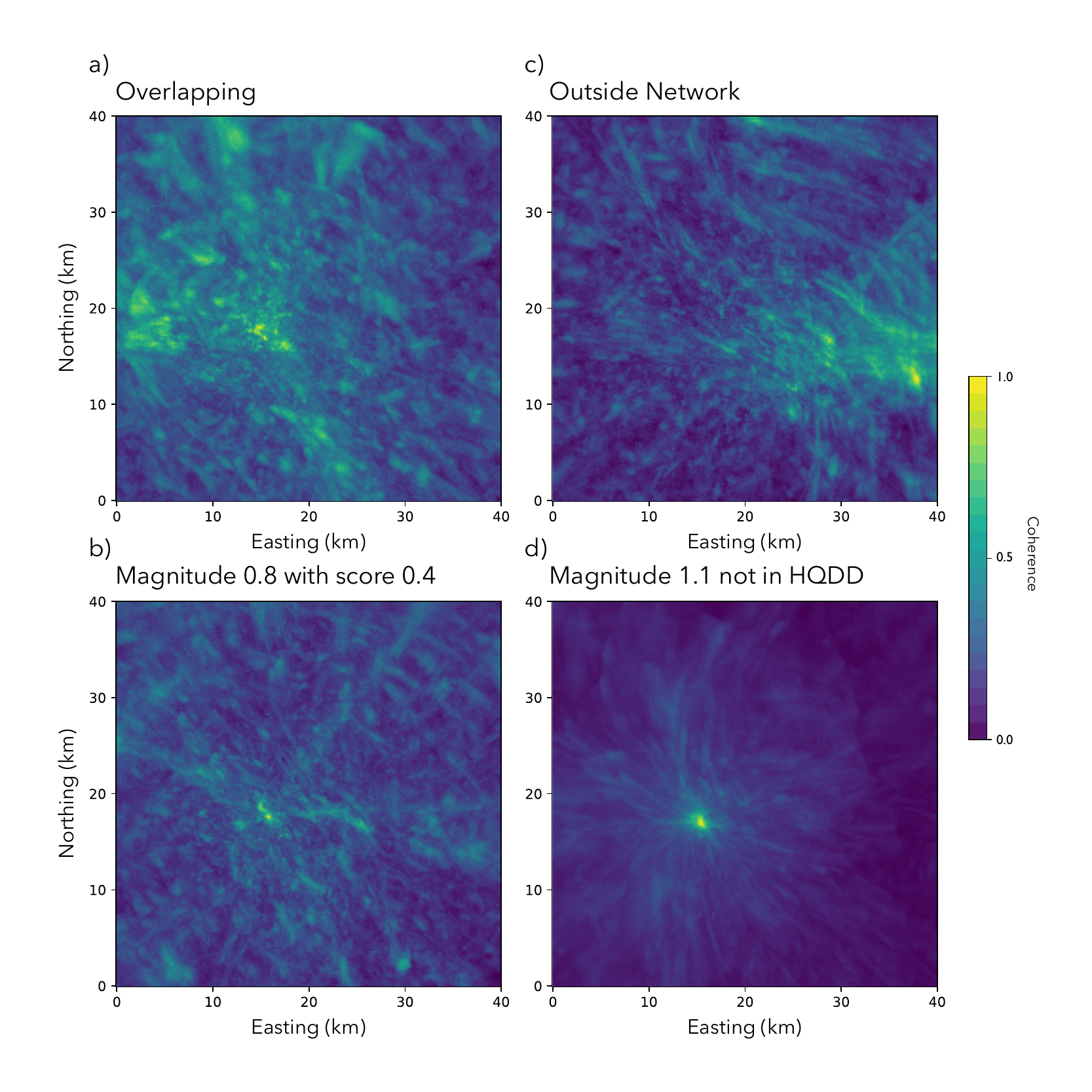}
\caption{Coherence matrices associated to: a) overlapping events; b) event outside the network; c) an event of magnitude 8 that was misclassified as noise, but the processing window contains two events with inter-event time less than 5 s) waveforms; d) Event of magnitude 1.1 that was correctly classified as event but that is not included in the HQDD catalogue.}
\label{fig9}
\end{figure}

It is important to note that all five events with magnitudes above 0.5 have classification scores greater than 0.1. Overall, 100 misclassified events (earthquakes classified as noise) exhibit classification scores above this threshold, while only 37 (out of 137) have scores lower than 0.1 (below the initial magnitude of completeness). Classification scores higher than 0.1 may indicate weak events that still display event-related patterns in the coherence images, although they remain similar to noise. A sample of coherence matrices for different event types is shown in Figure \ref{fig9}. These images display patterns that allow the classifier to identify the following cases as noise instances (Figure \ref{fig9}.a,b and c): overlapping earthquakes (Figure \ref{fig9}.a); events occurring outside the network (Figure \ref{fig9}.b). Figure \ref{fig9}.d shows a magnitude 1.1 event not present in the HQDD catalog, showing the capability of our approach to improve the seismic catalog. Overlapping or multiple events with short inter-event times (on the order of 5 s) introduce spurious effects in the coherence matrices that reduce classification accuracy. For this reason, we recommend manually checking the few events above the magnitude of completeness that are classified as noise.

\begin{figure}[ht]
\centering
\includegraphics[width=0.8\linewidth]{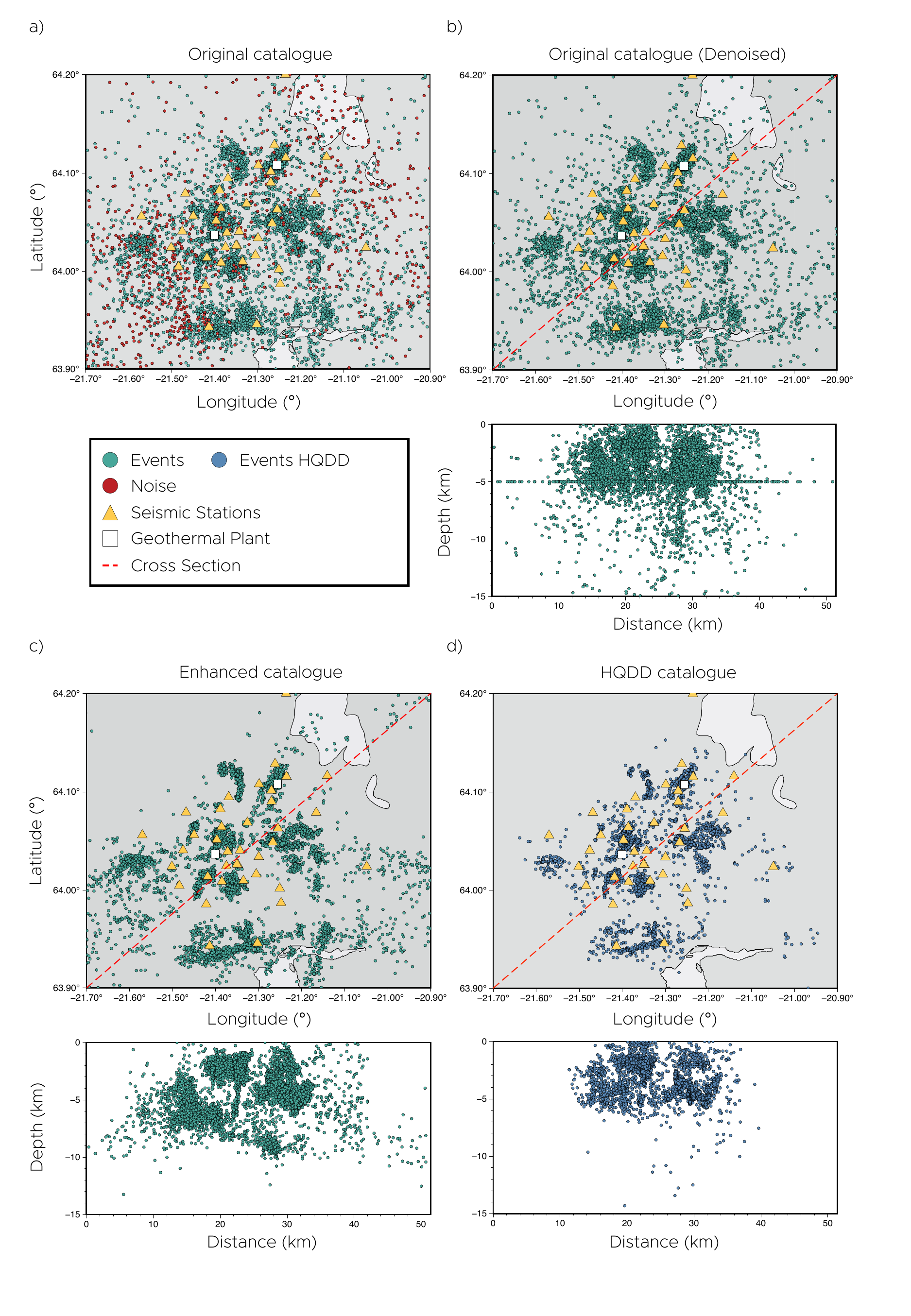}
\caption{Comparison between the original (a) and denoised seismic catalogues (b) and comparison of the spatial distribution of the enhanced catalogue (c) and the reference one (HQDD) (d). The enhanced catalogue includes events classified as real seismic events by the CNN and relocated using waveform stacking, showing a better resolution than HQDD catalogue, especially at depth, and increasing the number of events from 8500 to 11372.}
\label{fig10}
\end{figure}

Our method successfully recognizes as noise atypical seismic events occurring outside the area of interest, but that have been wrongly located within the LQ catalogue. Since the magnitude of these events may be relevant, they pollute the seismic catalog by reducing their usefulness for statistical seismology applications. For instance, Figure (E.1) in the electornic supplement, shows a low-frequency event of magnitude 2.5 that has been listed in the orignal LQ catalog and that has been correctly identified as noise by our approach, demonstrating the ability of the CNN to identify atypical events that we consider noise. Figure \ref{fig10} shows the comparison between the original catalog and the denoised one (Figures \ref{fig10}.a and \ref{fig10}.b), from which the 1002 noise-classified instances (red dots) have been removed (i.e.we keep only the misclassified events with classification score higher than 0.1). We observe that most of these noise-classified events occur outside the seismicity clusters. Since the application of waveform stacking methods yields to noise robust seismic event locations (\cite{Grigoli2013GRL}), our framework simultaneously cleans the catalog by removing noise and improves the location resolution of the seismicity, resulting in an enhanced catalog. Figure \ref{fig10} shows that our final catalog, comprising 11372 events, is in strong agreement with the high-quality double-difference (HQDD) reference catalog produced during the COSEISMIQ project (Figures \ref{fig10}.c and \ref{fig10}.d). It is important to note that events in the original LQ catalog are now more accurately relocated using our coherence-based method. While the HQDD catalog contains 8500 events, our enhanced catalog includes 11372 events, providing a more comprehensive characterization of the seismicity in the area, including seismic clusters that are missing in the HQDD catalog (Figure \ref{fig10}.d). 

Due to the lack of a local magnitude relationship specifically calibrated for the hengill area, and considering the short distances between stations and seismic sources (in many cases less than 5 km) the magnitude computation of small events is problematic. For this reason, we recalculated the magnitudes using a relative approach. We first applied the DBSCAN (Easter et al. 1991) clustering to the relocated seismicity catalog, then for each cluster we selected a representative master event with magnitudes ranging between ML 1.8 and 2.0. The events that are not initially assigned to any cluster (i.e. the ones that the DBSCAN considers noise) are subsequently associated with the nearest cluster centroid (Figure E.X). For each master event, the magnitude was calculated using the same approach proposed in Grigoli et al. (2022).  Noisy or unreliable stations were excluded from the computation. The final magnitude of the master event was obtained as the median of the single-station magnitudes, reducing the influence of outliers. The magnitudes of the other events within the same cluster were then estimated by comparing their amplitudes with those of the master event.
The relative magnitude was computed using the relation:
\begin{equation}
    M_{T}=M_{R}+\log{\left(\frac{A_{T}}{A_{R}}\right)}
\end{equation}
where $M_{R}$ and $M_{T}$ are, respectively, the magnitudes of the reference and target events, while $A_{R}$ and $A_{T}$ are the maximum amplitudes of the reference and target event measured at the same station (taken over the horizontal components). For each cluster, the same set of stations used for the calculation of the magnitude of the master event was retained, ensuring homogeneous observational conditions and minimizing possible biases related to station–source geometry.

\section*{Discussions and Conclusions}
Our results demonstrate that classifying coherence matrices with convolutional neural networks (CNNs) trained completely on synthetic data is an effective strategy for cleaning and enhancing seismic catalogues, particularly when dealing with recently deployed seismic monitoring network that lack of real data for training. By transforming raw seismic waveforms into coherence matrices, we overcome the limitations arising from the lack of similarity between synthetics and real data. Our workflow not only removes low-quality or spurious detections but also recovers events characterized by low quality locations, thereby enriching the location resolution of the final catalog. The enhanced catalog improves seismotectonic interpretations, supports more accurate statistical analyses (e.g., b-value estimation or its spatio-temporal evolution), and contributes to more reliable seismic hazard assessments studies. Importantly, the method is robust against noise and capable of identifying instances that conventional algorithms might miss or misclassify, such as low-frequency or off network events. For example, Figure (E.2) in the electronic supplement, shows a low-frequency M 2.5 event (probably of volcanic origin), that our method confidently classified as noise, underscoring its potential to improve catalog cleanliness and reliability. To minimize false positives, a more objective approach to identifying potential misclassifications is to consider only the misclassified noise-instances above the magnitude of completeness of the initial catalog. Applying this criterion, we retained in the final catalog only the five events misclassified as noise that exceeded this threshold. We also found that the method successfully flags atypical seismic events located outside the area of interest but initially included within the LQ catalogue, thereby reducing contamination from mislocated events. Beyond event classification and relocation, our framework is an approach for automating data quality analysis, a critical step for ensuring the reliability of downstream seismological products. Changes in b-values, like many statistical seismology parameters, are highly susceptible to artifacts in earthquake catalogs, including spatio-temporal variations in completeness (\cite{Herr_2022}), shifts or stretches of the magnitude scale, magnitude-dependent location uncertainties, changes in network processing procedures, or differences in magnitude types. Even novel large ML-based seismic catalogs are not immune to these issues. In this context our framework helps to fill this gap. Applied to the COSEISMIQ dataset, this coherence-based deep learning framework proves versatile, computationally efficient, and well-suited for both retrospective catalogue enhancement and near-real-time monitoring when combined with seismic event detection routines.

\section*{Data and Resources}
The continuous seismic data used in this study are available at the Swiss Seismological Service (SED) data center through network code: VI, OR and 2C (\url{http://eida.ethz.ch/fdsnws/dataselect/1/}, \cite{grigoli2022}). The obtained catalogs can be found in the supplemental materials. The earthquake catalog used for comparison are available at \cite{grigoli2022}, while the one compiled here can be found in the following repository (zenodo after the publication). The codes (NAMAZU), the related information to reproduce these results are available on github \url{https://github.com/XXXX}.

\section*{Author Contribution}
Conceptualization: MDS, FG. Methodology: MDS, FG. Software: MDS, FG, SG. Formal Analysis: MDS, FG, CR. Visualization: FG. Writing - Original draft: MDS, FG. Writing MDS, CR, FG - Review and Editing: MDS, FG, CR, SG, GR, BE.

\section*{Acknowledgments}
FG and BE thank the JPSP program, Fellowship ID S24079. EB was supported by the HORIZON EU project GEOHEAT (Grant Agreement 101147571). CR and SG have been funded by the Cofund CETP project GEOTWINS (Grant Agreement 101069750). LG was supported by the ERC project FARIA (Grant Agreement 101200403).


\begin{thebibliography}{}

\bibitem[{\textit{Abadi et~al.}}(2015)]{tensorflow}
Abadi, M., Agarwal, A., Barham, P., Brevdo, E., .. Zheng, X. (2015)
TensorFlow: Large-scale machine learning on heterogeneous systems,
2015. \url{tensorflow.org}

\bibitem[{\textit{Allen}}(1982)]{Allen1982} Allen, R. (1982). Automatic phase pickers: Their present use and future prospects. Bulletin of the Seismological Society of America, 72(6B), S225-S242.

\bibitem[{\textit{Arrowsmith et~al.}}(2022)]{arrowsmith_big_2022}
 Arrowsmith,S. J., Trugman,  D. T., MacCarthy,  J.,  Bergen, K. J.,  Lumley,D.,  Magnani, M. B. (2022) Big Data Seismology. Reviews of Geophysics, 60 (2): e2021RG000769

\bibitem[{\textit{Beyreuther et~al.}}(2010)]{Bey2010obspy} Beyreuther, M., Barsch, R., Krischer, L., Megies, T., Behr, Y., and Wassermann, J. (2010). ObsPy: A Python toolbox for seismology. Seismological Research Letters, 81(3), 530-533.


\bibitem[{\textit{Bergen et~al.}}(2019)]{bergen_machine_2019}
 Bergen K.J., Johnson , P.A., V. de Hoop, M., Beroza, G.C. (2019) Machine learning for data-driven discovery in solid Earth geoscience. Science, 363: eaau0323

\bibitem[{\textit{Chiaraluce et~al.}}(2022)]{chiaraluce2022} Chiaraluce, L., Michele, M., Waldhauser, F. et al. A comprehensive suite of earthquake catalogues for the 2016-2017 Central Italy seismic sequence. Sci Data 9, 710 (2022). https://doi.org/10.1038/s41597-022-01827-z

\bibitem[{\textit{Chollet et~al.}}(2015)]{keras}
Chollet, F. (2015), \url{https://keras.io}

\bibitem[{\textit{Fuchs et~al.}}(1971)]{Fuchs_1971}
Fuchs K. Müller G. (1971). Computation of synthetic seismograms with the reflectivity method and comparison with observations, Geophysical Journal International, 23 (4): 417- 433.

\bibitem[{\textit{Géron et~al.}}(2019)]{hands_on}
Géron, A. (2019). Hands-on machine learning with Scikit-Learn, Keras and TensorFlow: concepts, tools, and techniques to build intelligent systems (2nd ed.). O’Reilly

\bibitem[{\textit{Gharti et~al.}}(2010)]{Gharti2010} Gharti, H. N., Oye, V., Roth, M., and Kühn, D. (2010). Automated microearthquake location using envelope stacking and robust global optimization. Geophysics, 75(4), MA27-MA46.

\bibitem[{\textit{Grigoli et~al.}}(2013)]{Grigoli2013GRL} Grigoli, F., Cesca, S., Vassallo, M., and Dahm, T. (2013). Automated seismic event location by travel‐time stacking: An application to mining induced seismicity. Seismological Research Letters, 84(4), 666-677.

\bibitem[{\textit{Grigoli et~al.}}(2022)]{grigoli2022} Grigoli, F., Clinton, J., Diehl, T., Kaestil, P., Scarabello, L., Agustsdottir, T., ... and Dahm, T. (2022). Monitoring microseismicity in the Hengill Geothermal Field, Iceland. Submitted.

\bibitem[{\textit{Grigoli et~al.}}(2018)]{Grigoli2018GJI} Grigoli, F., Scarabello, L., Böse, M., Weber, B., Wiemer, S., and Clinton, J. F. (2018). Pick-and waveform-based techniques for real-time detection of induced seismicity. Geophysical Journal International, 213(2), 868-884.

  \bibitem[{\textit{He et~al.}}(2016)]{resnet}
He, K., Zhang, X., Ren, S., Sun, J.,  (2016) Deep Residual Learning for Image Recognition. 2016 IEEE Conference on Computer Vision and Pattern Recognition (CVPR), Las Vegas, NV, USA, 2016, pp. 770-778


 \bibitem[{\textit{Heimann et~al.}}(2019)]{Heimann_2019}
Heimann, S., Vasyura-Bathke, H., Sudhaus, H., Isken, M.P., Kriegerowski, M., Steinberg, A., Dahm, T.  (2019) A Python framework for efficient use of pre-computed Green's functions in seismological and other physical forward and inverse source problems. Solid Earth, 10(6):1921–1935 

\bibitem[{\textit{Helmholtz Centre Potsdam}}(2008)]{Helmholtz2008SDS} Helmholtz Centre Potsdam GFZ German Research Centre for Geosciences and gempa GmbH (2008). The SeisComP seismological software package. GFZ Data Services. doi:10.5880/GFZ.2.4.2020.003. 


\bibitem[{\textit{ Herrmann and Marzocchi}}(2020)]{Herr_Marz_2020}
Herrmann, M., and Marzocchi, W. (2020)  Inconsistencies and Lurking Pitfalls in the Magnitude–Frequency Distribution of High‐Resolution Earthquake Catalogs Available. Seismological Research Letters, 92 (2A): 909–922


\bibitem[{\textit{ Herrmann et~al.}}(2022)]{Herr_2022}
Herrmann, M., Piegari, E.,  Marzocchi , W. (2022) Revealing the spatiotemporal complexity of the magnitude distribution and b-value during an earthquake sequence. Nat Commun 13: 5087

\bibitem[{\textit{Hunter}}(2007)]{Hunter} Hunter, J. D. (2007). Matplotlib: A 2D graphics environment. Computing in science \& engineering, 9(03), 90-95.

\bibitem[{\textit{Kao and Shan}}(2004)]{KaoSSA} Kao, H., and Shan, S. J. (2004). The source-scanning algorithm: Mapping the distribution of seismic sources in time and space. Geophysical Journal International, 157(2), 589-594.

\bibitem[{\textit{Kingma and Ba}}(2015)]{adam}
Kingma, D.P., and Ba, J. (2015) Adam: A Method for Stochastic Optimization. Conference Paper at the 3rd International Conference for Learning Representations, San Diego.

\bibitem[{\textit{Kong et~al.}}(2019)]{Kong2019} Kong, Q., Trugman, D.T., Ross,  Z.E., Bianco, M.J., Meade,  B.J., and Gerstoft,  P. (2019). Machine learning in seismology: Turning data into insights. Seismological Research Letters, 90(1), 3–14.

\bibitem[{\textit{Langet et~al.}}(2014)]{Langet2014} Langet, N., Maggi, A., Michelini, A., and Brenguier, F. (2014). Continuous kurtosis‐based migration for seismic event detection and location, with application to Piton de la Fournaise Volcano, La Reunion. Bulletin of the Seismological Society of America, 104(1), 229-246.

  \bibitem[{\textit{LeCun et~al.}}(2010)]{LeCun}
LeCun, Y., Kavukcuoglu, K., and Farabet, C.(2010) Convolutional networks and applications in vision. Proceedings of 2010 IEEE International Symposium on Circuits and Systems, Paris, France, pp. 253-256


  \bibitem[{\textit{Li et~al.}}(2020)]{Li_2020}
Li, L., Tan, J., Schwarz, B., Staněk, F., Poiata, N., Shi, P.,  Diekmann, L., Eisner, L., GajewskiStaněk, D. (2020) Recent Advances and Challenges of Waveform-Based Seismic Location Methods at Multiple Scales. Reviews of Geophysics, 58(1):e2019RG000667


\bibitem[{\textit{Liao et~al.}}(2021)]{Liao2021deve} Liao S., Zhang H., Fan L., Li B., Huang L., Fang L., and Qin M. (2021). Development of a real-time intelligent seismic processing system and its application in the 2021 Yunnan Yangbi MS6.4 earthquake. Chinese Journal of Geophysics, 64(10): 3632-3645.

\bibitem[{\textit{Lim et~al.}}(2025)]{Lim2025} Lim C., Lapins S., Segou M., Werner M., Deep learning phase pickers: how well can existing models detect hydraulic-fracturing induced microseismicity from a borehole array?, Geophysical Journal International, Volume 240, Issue 1, January 2025, Pages 535–549.

\bibitem[{\textit{Liu et~al.}}(2020)]{Liu2020ridg} Liu, M., Zhang, M., Zhu, W., Ellsworth, W. L., and Li, H. (2020). Rapid characterization of the July 2019 Ridgecrest, California, earthquake sequence from raw seismic data using machine‐learning phase picker. Geophysical Research Letters, 47(4), e2019GL086189.

\bibitem[{\textit{ Lomax et~al.}}(2024)]{DKPN}
Lomax, A., Bagagli, M., Gaviano, S., Cianetti, S., Jozinović, D., Michelini, A., Zerafa, C., Giunchi, C. (2024)  Effects on a Deep-Learning, Seismic Arrival-Time Picker of Domain-Knowledge Based Preprocessing of Input Seismograms. Seismica

\bibitem[{\textit{Mancini et~al.}}(2022)]{mancini} Mancini, S., Segou, M., Werner, M. J., Parsons, T., Beroza, G., \& Chiaraluce, L. (2022). On the use of high-resolution and deep-learning seismic catalogs for short-term earthquake forecasts: Potential benefits and current limitations. Journal of Geophysical Research: Solid Earth, 127, e2022JB025202. https://doi.org/10.1029/2022JB025202

\bibitem[{\textit{Majstorović et~al.}}(2021)]{JosipaJGR2021} Majstorović, J., Giffard‐Roisin, S., and Poli, P. (2021). Designing Convolutional Neural Network Pipeline for Near‐Fault Earthquake Catalog Extension Using Single‐Station Waveforms. Journal of Geophysical Research: Solid Earth, 126(7), e2020JB021566.

\bibitem[{\textit{ McBrearty et~al.}}(2019)]{McBrearty_2019}
 McBrearty, I.W.,  Delorey, A.A., Johnson, P.A. (2019)  Pairwise Association of Seismic Arrivals with Convolutional Neural Networks. Seismological Research Letters, 90 (2A): 503–509


\bibitem[{\textit{ McBrearty and Beroza}}(2023)]{genie}
 McBrearty, I.W., and  Beroza, G.C. (2023) Earthquake Phase Association with Graph Neural Networks Available. Bulletin of the Seismological Society of America, 113 (2): 524–547

\bibitem[{\textit{Mousavi et~al.}}(2019)]{CRED} Mousavi, S. M.,  Zhu, W.,  Sheng, Y., Beroza, G.C. (2019). CRED: A deep residual network of convolutional and recurrent units for earthquake signal detection. Scientific Reports, 9, 10267

\bibitem[{\textit{Mousavi et~al.}}(2020)]{eqtransformer} Mousavi, S. M., Ellsworth, W. L., Zhu, W., Chuang, L. Y., and Beroza, G. C. (2020). Earthquake transformer—an attentive deep-learning model for simultaneous earthquake detection and phase picking. Nature communications, 11(1), 1-12.

\bibitem[{\textit{Mousavi and Beroza}}(2022)]{Mous-Ber2022} Mousavi, S.M., and  Beroza,G.C. (2022) Deep-learning seismology. Science, 377(6607), eabm4470

\bibitem[{\textit{Munchmeyer et~al.}}(2022)]{Munchmeyer2022}
Münchmeyer, J.,  Woollam, J.,  Rietbrock, A., Tilmann, F.,  Lange, D.,  Bornstein, T.,  Diehl, T., Giunchi, C., Haslinger, F., Jozinović, D.,  Michelini, A., Saul, J., Soto , H. (2022) Which Picker Fits My Data? A Quantitative Evaluation of Deep Learning Based Seismic Pickers. JGR Solid Earth, 127(1): e2021JB023499 

  \bibitem[{\textit{ Münchmeyer}}(2024)]{PyOcto}
Münchmeyer, J.  (2024)  PyOcto: A high-throughput seismic phase associator. Seismica

\bibitem[{\textit{Nakata and Beroza}}(2016)]{NakataRTM} Nakata, N., and Beroza, G. C. (2016). Reverse time migration for microseismic sources using the geometric mean as an imaging condition. Geophysics, 81(2), KS51-KS60.


\bibitem[{\textit{Nooshiri et~al.}}(2021)]{Nooshiri_2021}
Nooshiri ,N., Bean , C.J.,  Dahm , T.,  Grigoli , F., Kristjánsdóttir , S., Obermann , A., Wiemer, S. (2021)
A multibranch, multitarget neural network for rapid point-source inversion in a microseismic environment: examples from the Hengill Geothermal Field, Iceland. Geophysical Journal International. 229(2): 999–1016

\bibitem[{\textit{Olivieri and Clinton}}(2012)]{Olivieri} Olivieri, M. and Clinton, J., 2012. An almost fair comparison between Earth- worm and SeisComp3, Seismol. Res. Lett., 83(4), 720–727.

\bibitem[{\textit{Park et~al.}}(2021)]{park2021} Park, Y., Beroza, G. C., and Ellsworth, W. L. (2021). A Deep Earthquake Catalog for Oklahoma and Southern Kansas Reveals Extensive Basement Fault Networks. Earth and Space Science Open Archive. https://doi.org/10.1002/essoar.10508504.1

  \bibitem[{\textit{Poiata et~al.}}(2016)]{bbtrack}
Poiata, N.,  Satriano, C.,  Vilotte, J.P., Bernard, P.,  Obara, K. (2016) Multiband array detection and location of seismic sources recorded by dense seismic networks. Geophysical Journal International,205(3): 1548–1573

\bibitem[{\textit{Reykjavik Energy}}(2016)]{Reykjavik2016} Reykjavik Energy (Iceland). (2016). OR - Reykjavik Energy [Data set]. International Federation of Digital Seismograph Networks. https://doi.org/10.7914/SN/OR

\bibitem[{\textit{Ross et~al.}}(2018)]{GPD} Ross, Z. E., Meier, M. A., Hauksson, E., and Heaton, T. H. (2018). Generalized seismic phase detection with deep learning. Bulletin of the Seismological Society of America, 108(5A), 2894-2901.

\bibitem[{\textit{Ross et~al.}}(2019)]{phaselink_2019}
 Ross, Z.E., Yue, Y., Meier, MA., Hauksson, E., Heaton, T.H. (2019)
PhaseLink: A Deep Learning Approach to Seismic Phase Association. JGR Solid Earth, 124(1): 856-869

\bibitem[{\textit{Rossi et~al.}}(2020)]{Rossi2020full} Rossi, C., Grigoli, F., Cesca, S., Heimann, S., Gasperini, P., Hjörleifsdóttir, V., Dahm, T., Bean, C.J., Wiemer, S., Scarabello, L. and Nooshiri, N. (2020). Full-Waveform based methods for Microseismic Monitoring Operations: an Application to Natural and Induced Seismicity in the Hengill Geothermal Area, Iceland. Advances in Geosciences, 54, 129-136.

\bibitem[{\textit{Scarabello }}(2021)]{RTDD} Scarabello, L. (2021). swiss-seismological-service/scrtdd: v1.6.1 (v1.6.1). Zenodo. https://doi.org/10.5281/zenodo.5337361

\bibitem[{\textit{SED at ETH}}(2018)]{sed2018} Swiss Seismological Service (SED) at ETH Zurich. (2018). COSEISMIQ - COntrol SEISmicity and Manage Induced earthQuakes. ETH Zurich. https://doi.org/10.12686/sed/networks/2c

\bibitem[{\textit{Shi et~al.}}(2019a)]{shimcm1} Shi, P., Angus, D., Rost, S., Nowacki, A., and Yuan, S. (2019). Automated seismic waveform location using multichannel coherency migration (MCM)–I: theory. Geophysical Journal International, 216(3), 1842-1866.

\bibitem[{\textit{Shi et~al.}}(2019b)]{shimcm2} Shi, P., Nowacki, A., Rost, S., and Angus, D. (2019). Automated seismic waveform location using Multichannel Coherency Migration (MCM)—II. Application to induced and volcano-tectonic seismicity. Geophysical Journal International, 216(3), 1608-1632.

\bibitem[{\textit{Shi et~al.}}(2021)]{shi2021} Shi, P., Seydoux, L., and Poli, P. (2021). Unsupervised learning of seismic wavefield features: clustering continuous array seismic data during the 2009 L'Aquila earthquake. Journal of Geophysical Research: Solid Earth, 126(1), e2020JB020506.

\bibitem[{\textit{Shi et~al.}}(2022)]{malmi_2022}
 Shi, P., Grigoli, F., Lanza, F., Beroza, G.C., Scarabello, L., Wiemer, S. (2022)
 MALMI: An Automated Earthquake Detection and Location Workflow Based on Machine Learning and Waveform Migration Available. Seismological Research Letters, 93 (5): 2467–2483

  \bibitem[{\textit{ Si et~al.}}(2024)]{Si_2024}
 Si, X., Wu, X., Li, Z., Wang, S., Zhu, J.  (2024)  An all-in-one seismic phase picking, location, and association network for multi-task multi-station earthquake monitoring. Commun Earth Environ 5, 22

 \bibitem[{\textit{ Soto and Schurr}}(2021)]{DPP}
Soto, H., Schurr, B. (2021)  DeepPhasePick: a method for detecting and picking seismic phases from local earthquakes based on highly optimized convolutional and recurrent deep neural networks. Geophysical Journal International, 227(2): 1268–1294


  \bibitem[{\textit{Staněk et~al.}}(2015)]{stanek}
Staněk, F., Anikiev, D., Valenta, J., Eisner, L.(2015) Semblance for microseismic event detection. Geophysical Journal International, 201(3):1362–1369

\bibitem[{\textit{Tan et~al.}}(2021)]{tan2021} Tan, Y. J., Waldhauser, F., Ellsworth, W. L., Zhang, M., Zhu, W., Michele, M., ... and Segou, M. (2021). Machine‐Learning‐Based High‐Resolution Earthquake Catalog Reveals How Complex Fault Structures Were Activated during the 2016–2017 Central Italy Sequence. The Seismic Record, 1(1), 11-19.

\bibitem[{\textit{Tryggvason et~al.}}(2002)]{Tryggvason}
Tryggvason, A. , Rögnvaldsson , S.T., Flóvenz , O.G. (2002) Three-dimensional imaging of the P- and S-wave velocity structure and earthquake locations beneath Southwest Iceland. Geophysical Journal International, 151(3):848–866

\bibitem[{\textit{Utsu}}(1961)]{Utsu1961} Utsu, T. (1961). A statistical study on the occurrence of aftershocks. Geophys. Mag., 30, 521-605.

\bibitem[{\textit{Wessel et~al.}}(2019)]{GMT6} Wessel, P., Luis, J. F., Uieda, L., Scharroo, R., Wobbe, F., Smith, W. H. F., and Tian, D. (2019). The Generic Mapping Tools version 6. Geochemistry, Geophysics, Geosystems, 20, 5556–5564. https://doi.org/10.1029/2019GC008515

\bibitem[{\textit{Waldhauser}}(2009)]{Waldhauser2009} Waldhauser, F. (2009). Near-real-time double-difference event location using long-term seismic archives, with application to Northern California. Bulletin of the Seismological Society of America, 99(5), 2736-2748.

\bibitem[{\textit{Waldhauser and Ellsworth}}(2000)]{Waldhauserdd} Waldhauser, F. and Ellsworth, W. L. (2000). A double-difference earthquake location algorithm: Method and application to the northern Hayward fault, California. Bulletin of the Seismological Society of America, 90(6), 1353-1368.

\bibitem[{\textit{Wang}}(1999)]{Wang_1999}
Wang, R. (1999) A simple orthonormalization method for stable and efficient computation of Green's functions.  Bulletin of the Seismological Society of America, 89 (3): 733–741

\bibitem[{\textit{Werner and Saenger}}(2018)]{claudia2018} Werner, C., and Saenger, E. H. (2018). Obtaining reliable source locations with time reverse imaging: limits to array design, velocity models and signal-to-noise ratios. Solid Earth, 9(6), 1487-1505.

\bibitem[{\textit{Withers et~al.}}(1998)]{Withers1998} Withers, M., Aster, R., Young, C., Beiriger, J., Harris, M., Moore, S., and Trujillo, J. (1998). A comparison of select trigger algorithms for automated global seismic phase and event detection. Bulletin of the Seismological Society of America, 88(1), 95-106.

\bibitem[{\textit{Willacy et~al.}}(2019)]{Willacy2019} Willacy, C., van Dedem, E., Minisini, S., Li, J., Blokland, J. W., Das, I., and Droujinine, A. (2019). Full-waveform event location and moment tensor inversion for induced seismicity. Geophysics, 84(2), KS39-KS57.


\bibitem[{\textit{ Woollam et~al.}}(2019)]{Woollam_2019}
Woollam, J., Rietbrock, A., Bueno, A., and De Angelis, S. (2019)  Convolutional neural network for seismic phase classification, performance demonstration over a local seismic network. Seismological
Research Letters, 90:491-502


\bibitem[{\textit{ Woollam et~al.}}(2020)]{Woollam_2020}
Woollam, J., Rietbrock, A., Leitloff, J., and Hinz, S. (2020)  HEX: Hyperbolic Event EX-tractor, a seismic phase associator for highly active seismic regions. Seismological
Research Letters, 91 (7):2769–78


\bibitem[{\textit{Woollam et~al.}}(2022)]{seisbench}
 Woollam, J., Münchmeyer, J., Tilmann, F., Rietbrock, A., Lange, D., Bornstein, T., Diehl, T.,  Giunchi, C.,  Haslinger, F., Jozinović, D.,  Michelini, A.,  Saul, J., Soto, H. (2022) SeisBench—A Toolbox for Machine Learning in Seismology. Seismological Research Letters, 93(3):1695–1709


 \bibitem[{\textit{ Yu and Wang}}(2022)]{LPPN}
Yu, Z., Wang, W. (2022)  LPPN: A lightweight network for fast phase picking. Seismological
Research Letters, 93: 2834-2846


\bibitem[{\textit{Zhang et~al.}}(2022)]{zhang2022locflow} Zhang, M., Liu, M., Tian, T., Wang, R., and Zhu, W. (2022). LOC-FLOW: An end-to-end machine-learning-based high-precision earthquake location workflow. Seismological Research Letters.

 \bibitem[{\textit{Zhao et~al.}}(2024)]{Zhao_2024}
Zhao, X., Wang, L., Zhang, Y., Han, X.,  Deveci, M., and  Parmar, M.  (2024) A review of convolutional neural networks in computer vision. Artif Intell Rev, 57(99) 

\bibitem[{\textit{Zhou et~al.}}(2021)]{zhoupalm} Zhou, Y., Yue, H., Fang, L., Zhou, S., Zhao, L., and Ghosh, A. (2021). An earthquake detection and location architecture for continuous seismograms: phase picking, association, location, and matched filter (PALM). Seismological Research Letters, 93(1), 413-425.

\bibitem[{\textit{Zhou et~al.}}(2019)]{ppknet} Zhou, Y., Yue, H., Kong, Q., and Zhou, S. (2019). Hybrid event detection and phase‐picking algorithm using convolutional and recurrent neural networks. Seismological Research Letters, 90(3), 1079-1087.

\bibitem[{\textit{Zhu et~al.}}(2019)]{CPIC} Zhu, L., Peng, Z., McClellan, J., Li, C., Yao, D., Li, Z., and Fang, L. (2019). Deep learning for seismic phase detection and picking in the aftershock zone of 2008 Mw7. 9 Wenchuan Earthquake. Physics of the Earth and Planetary Interiors, 293, 106261.

\bibitem[{\textit{Zhu and Beroza.}}(2019)]{phasenet} Zhu, W., and Beroza, G. C. (2019). PhaseNet: a deep-neural-network-based seismic arrival-time picking method. Geophysical Journal International, 216(1), 261-273.

\bibitem[{\textit{ Zhu et~al.}}(2022)]{gamma}
 Zhu, W.,  McBrearty, I.W., Mostafa , S., Mousavi, Ellsworth, W.L., Beroza, G. C. (2022)  Earthquake Phase Association Using a Bayesian Gaussian Mixture Model. JGR Solid Earth, 127(5): e2021JB023249

\end{thebibliography}
\end{document}